\documentclass[11pt,twoside]{article}
\usepackage{asp2010}
\usepackage{graphicx}
\usepackage{natbib}

\resetcounters

\begin{document}

\title{Photometric Survey to Search for Field sdO Pulsators}
\author{Christopher B. Johnson$^1$, E. M. Green$^2$, S. C. Wallace$^3$, C. J. O'Malley$^3$, \\H. Amaya$^3$, L. Biddle$^3$, G. Fontaine$^4$
\affil{$^1$Department of Physics and Astronomy, Louisiana State University, Baton Rouge, LA, 70803, USA}
\affil{$^2$Steward Observatory, University of Arizona, Tucson, AZ, 85721, USA}
\affil{$^3$Department of Astronomy, University of Arizona, Tucson, AZ, 85721, USA}}
\affil{$^4$D\'{e}partement de Physique, Universit\'{e} de Montr\'{e}al, Montr\'{e}al, Qu\'{e}bec, Canada, H3C 3J7}

\begin{abstract} We present the results of a campaign to search for 
subdwarf O (sdO) star pulsators among bright field stars. 
 The motivation for this 
project is the 
recent discovery by \citet{2011ApJ...737L..27R}, of four rapidly pulsating sdO 
stars in the globular cluster $\omega$ Cen, with T$_{\rm eff}$ near 50,000 K, 5.4 $<$ log ${g}$ $<$ 6.0,  and hydrogen-rich atmospheres. The only previously 
known sdO pulsator is significantly hotter at 68,500 K and 
log ${g}$ = 6.1. All of the sdO pulsators identified so far are fainter than $V\approx17.4$ and, thus, are poor candidates for an in-depth follow-up with asteroseismology. We therefore obtained high S/N light curves and spectroscopy for a number of field sdO stars to attempt to 
discover bright counterparts to these stars, particularly the $\omega$ Cen pulsators. 
Our primary sample consisted of 19 sdO stars with 
hydrogen-rich atmospheres, log $N$(He)$/N$(H) $<$ --1.0, effective 
temperatures in the range 40,000~K $<$ T$_{\rm eff}$ $<$ 67,000~K, and 
surface gravities 5.3 $<$ log ${g}$ $<$ 6.1. We also observed 17 additional helium-rich sdO stars with log $N$(He)$/N$(H) $>$ --0.1 and similar temperatures and gravities. 
To date, we have found no detectable pulsations at amplitudes above 0.08\% (4 times the mean noise level) in any of the 36 field sdO stars that we observed. The presence of pulsations in $\omega$ Cen sdO stars and their apparent absence in seemingly comparable field sdO stars is perplexing. While very suggestive, the significance of this result is difficult to assess more completely right now due to remaining uncertainties about the temperature width and purity of the $\omega$ Cen instability strip and the existence of any sdO pulsators with weaker amplitudes than the current detection limit in globular clusters.
\end{abstract}

\section{Introduction and sdO Target Selection} To date, six pulsating sdO stars are known. In 2006, \citet{2006MNRAS.371.1497W} made a serendipitous discovery of the first pulsating sdO star, J160043+074802.9, during a search for AM CVn stars. At least 10 different pulsation periods were identified in the range of 60--120 s. In a follow up study, \citet{2011ApJ...733..100L} derived the following parameters from a detailed spectral analysis: T$_{\rm eff}$ = 68,500 $\pm$ 1,770~K, log $g$ = 6.09 $\pm$ 0.07, and log $N$(He)$/N$(H) = --0.64 $\pm$ 0.05. In 2011, Randall et al.'s (2011) spectral analysis indicated that four newly identified, rapidly pulsating stars on the extreme horizontal branch (EHB) in $\omega$ Cen were sdO stars, the first multi-mode pulsating hot subdwarf stars ever found in a globular cluster (GC). They have H-rich atmospheres (--1.9 $<$ log $N$(He)$/N$(H) $<$ --1.2), T$_{\rm eff}$ between 48,000 and 52,000~K, log $g$ between 5.3 and 6.0, periods ranging from 84 to 124 seconds, and pulsation amplitudes of 2.7\% down to 0.5\%. A sixth pulsating He-sdO was found during a study of NGC\,2808 using far-UV observations by Brown et al. (2013) (this volume). The periods of all known sdO pulsators are consistent with p-mode pulsations.\\
\indent Homogenous parameters of our 36 target sdOs are summarized in Table 1, covering the ranges of 40,000~K $<$ T$_{\rm eff}$ $<$ 72,000 K and 5.1 $<$ log $g$ $<$ 6.3, with particular emphasis on effective temperatures near 50,000~K. More than half (19) of the sdO's in our sample have solar or less than solar abundances similar to those found in $\omega$ Cen. We also observed 17 He-sdOs because the latter are more numerous, especially near 50,000~K. Four of the He-sdO's (HZ\,1, PG\,0039+139, PG\,1427+196, and HS\,1707+6121) were previously observed by \citet{2007MNRAS.379.1123R}, in the first extensive search for sdO pulsators. Although \citet{2007MNRAS.379.1123R} considered nearly all of their 56 sdO targets to be non-variable, they listed both \\PG\,1427+196 and HS\,1707+6121 as very promising candidates on the basis of amplitude detection limits of $\sigma$ $<$ 1 mmag and apparent pulsation amplitudes greater than 3 to 4$\sigma$. We show our light curves and the resulting Fourier Transforms for these two stars in Figures 1 and 2, respectively.

\section{Observations and Data Reductions} Time series photometry was obtained with the Mont4k CCD on the Steward Observatory 1.55 m Mt. Bigelow telescope near Tucson, Arizona. The detector is a backside illuminated Fairchild 486 CCD with 4096 x 4097 15 $\mu$ pixels which was processed for high blue sensitivity. The Mont4k CCD is equipped with a suite of filters which include Bessell U, Harris BVR, Arizona I, as well as WFPC2 F555W, F606W, and F814W filters and a very wide band Schott 8612 filter. The pixel scale is 0.14''/pixel, yielding a field of view of 9.7' x 9.7'.\\
\indent We used the Schott 8612 filter for most of our light curves, except for a few very bright targets that were observed in B, V, or I filters to avoid saturation. The images were reduced using standard IRAF reduction tasks for bias subtraction and flat fielding. To generate the light curves, we used fluxes derived from aperture photometry for the target star relative to the average of a carefully selected set of field stars in the field of view of the image. The reference stars were chosen to have the most symmetric distribution about the target star as possible. The resulting relative flux was plotted against time to create all the light curves for the observing campaign.\\
\indent The Fourier Transforms of the light curves were calculated using the Period program by Starlink\footnote[1]{ http://www.starlink.rl.ac.uk/docs/sun167.htx/sun167.html} which is a time series analysis package and is available as an open source download. This program uses the Lomb-Scargle technique to search for the maximum peak in the power spectrum of a given data set.\\ 
\indent We also obtained homogeneous high S/N low resolution (9\AA) spectra for all of the sdOÕs in our sample using the B\&C spectrograph from the Bok 2.3 m telescope at Kitt Peak National Observatory. The spectra cover the wavelength range from 3625--6900 \AA. They were reduced using standard IRAF tasks. The Balmer and helium lines in the resulting flux-calibrated spectra were fit with a grid of NLTE H+He atmospheres to produce the parameters in Table 1.

\subsection{Light Curves} We present light curves in Figure 1 from 7 of our 87 nights of observation, including both H-rich and He-rich sdO's. This campaign encompassed almost two years of observations from May 2011 until April 2013. Each of these targets was observed for a minimum of 4--5 hours with  typical sampling times of 25--35 s. Some targets were observed for 4 or more hours on a single night,  while the observations for other targets were split between two or three closely spaced nights. The small sampling time allowed us to search for pulsations on the order of 60--600 s, which cover the regime of p-mode pulsations in all hot subdwarfs, sdO and sdB. By visual inspection, there are no obvious pulsations present in the data.  

\begin{figure}[Ht!]
\begin{center}
\includegraphics[scale = 0.6]{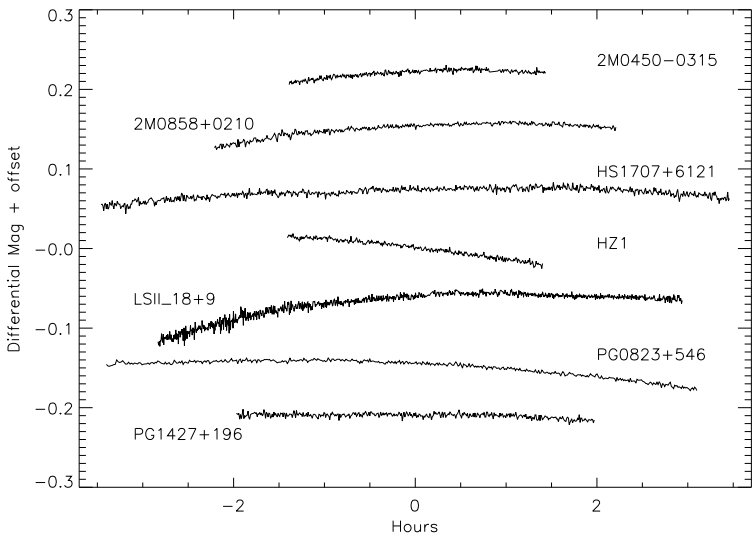}
\caption{Light curves of seven pulsating sdO candidates. The curves are plotted about the central time of observation for each night so that they could easily be stacked. The y-axis reflects the differential magnitude relative to the reference stars, vertically offset for clarity. The overall trends in brightness are due to differential refraction as a function of airmass between bluer target stars and the much redder reference stars. }
\label{default}
\end{center}
\end{figure}

%\newpage
\subsection{Fourier Transforms} We present Fourier transforms of two representative targets in Figure 2, PG\,1427+196 and HS\,1707+6121, which also happen to be the two best sdO pulsator candidates proposed by \citet{2007MNRAS.379.1123R}. The x-axis spans the entire region where we would expect to see peaks corresponding to the p-mode pulsations identified in all previously discovered sdO pulsators (not to mention all known sdB p-mode pulsators). The dashed line is four times the standard deviation in this region of the FT. The dotted line illustrates the smallest amplitude peak detected by \citet{2011ApJ...737L..27R} for similar T$_{\rm eff}$,  log $g$, and log $N$(He)$/N$(H). Out of 36 candidates, we found zero sdO's, either H-rich or He-rich, exhibiting pulsations above our 4$\sigma$ detection limit, which is typically about 0.08\% for each target. We find nothing remotely as large as the pulsation amplitudes found in \citet{2011ApJ...737L..27R}. In particular, we were unable to corroborate the results of \citet{2007MNRAS.379.1123R} for PG\,1427+196 and HS\,1707+6121 even though our observations were taken over much longer time intervals, 2--4 times as long in some instances. This is especially true for HS\,1707+6121 ($\sim$7 hours in this campaign compared to $\sim$2.88 hours in \citet{2007MNRAS.379.1123R}).   

\begin{figure}[Ht!]
\begin{center}
\includegraphics[width=2.5in,height=4in,angle=-90.0]{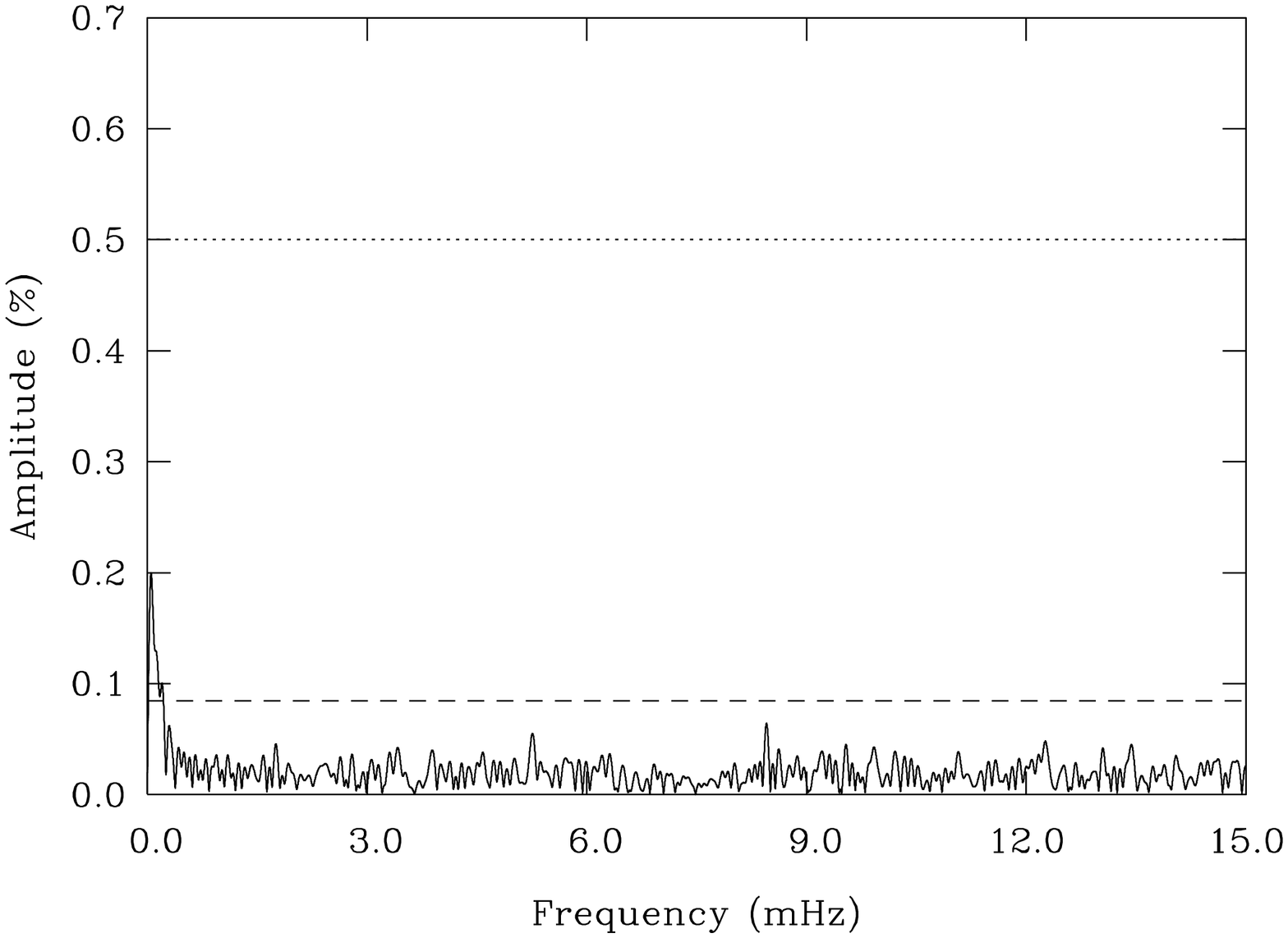}
\includegraphics[width=2.5in,height=4in,angle=-90.0]{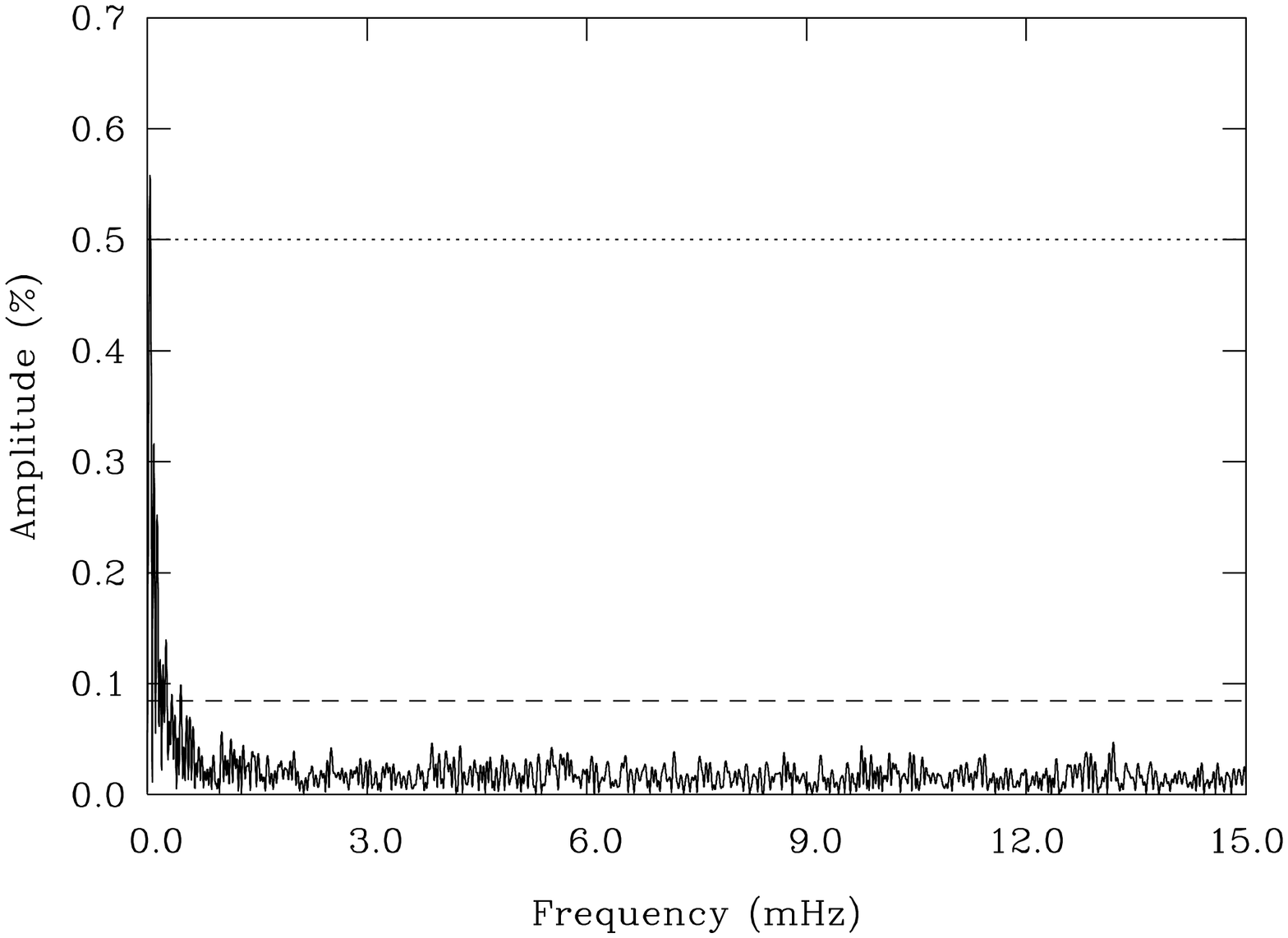}
\caption{The Fourier Transform of a single night's data for each of two representative sdO pulsator candidates, top: PG\,1427+196 and bottom: HS\,1707+6121. The dashed line is our 4$\sigma$ detection limit for each star and the dotted line is the smallest amplitude pulsation detected by \citet{2011ApJ...737L..27R} for the $\omega$ Cen pulsators.}
\label{default}
\end{center}
\end{figure}

\begin{table}[Ht!]
\caption{Pulsating sdO Candidate Data Table}
\centering
\begin{tabular}{c c c r}\\
\hline\hline 
object & T$_{\rm eff}$ & log $g$ & log $N$(He)$/N$(H) \\  [1 ex]% inserts table %heading
\hline
{\bf19 H-rich sdO stars}\\
\hline 

PG\,1610+519      & 40740 $\pm$  526 &  5.481 $\pm$ 0.063 & $-$2.989 $\pm$ 0.311\\
PG\,1000+408      & 40804 $\pm$  331 &  5.471 $\pm$ 0.040 & $-$2.724 $\pm$ 0.148\\
HS\,0252+1025     & 41573 $\pm$  625 &  5.204 $\pm$ 0.070 & $-$2.908 $\pm$ 0.285\\
PG\,0942-029      & 42300 $\pm$  304 &  6.194 $\pm$ 0.038 & $-$5.115 $\pm$ 0.065\\
2M0843+0200     & 42433 $\pm$  467 &  5.584 $\pm$ 0.045 & $-$3.005 $\pm$ 0.231\\
2M0858+0210     & 44566 $\pm$  442 &  6.007 $\pm$ 0.047 & $-$1.757 $\pm$ 0.062\\
2M0813+5257     & 45153 $\pm$  537 &  5.952 $\pm$ 0.053 & $-$1.919 $\pm$ 0.081\\
HS\,0222+2324     & 47390 $\pm$  828 &  5.797 $\pm$ 0.065 & $-$1.971 $\pm$ 0.052\\
FBS\,1715+424     & 49746 $\pm$  614 &  6.046 $\pm$ 0.044 & $-$3.296 $\pm$ 0.145\\
BD$-$11${\deg}$\,162       & 50638 $\pm$ 1323 &  5.583 $\pm$ 0.086 & $-$1.971 $\pm$ 0.129\\
2M0450-0315     & 51673 $\pm$  935 &  5.076 $\pm$ 0.069 & $-$1.123 $\pm$ 0.054\\
PG\,1506-052      & 52814 $\pm$ 1263 &  5.696 $\pm$ 0.063 & $-$1.972 $\pm$ 0.056\\
PG\,1409-103      & 53482 $\pm$ 1015 &  5.787 $\pm$ 0.069 & $-$2.046 $\pm$ 0.112\\
LSII+18\_9       & 54013 $\pm$ 1453 &  5.741 $\pm$ 0.071 & $-$1.901 $\pm$ 0.067\\
Ton835          & 55753 $\pm$ 1821 &  5.706 $\pm$ 0.062 & $-$2.135 $\pm$ 0.058\\
Feige67         & 56061 $\pm$ 1079 &  5.730 $\pm$ 0.048 & $-$1.811 $\pm$ 0.048\\
PG\,1708+602      & 58472 $\pm$ 1173 &  5.685 $\pm$ 0.062 & $-$0.991 $\pm$ 0.047\\
Ton357          & 60357 $\pm$ 1846 &  5.695 $\pm$ 0.064 & $-$1.840 $\pm$ 0.067\\
PG\,1102+499      & 67048 $\pm$ 4035 &  5.357 $\pm$ 0.106 & $-$1.190 $\pm$ 0.095\\
\hline
{\bf17 He-rich sdO stars}\\
\hline
HZ\,1             & 40663 $\pm$  423 &  5.477 $\pm$ 0.102 &  3.084 $\pm$ 0.521\\
BD+25${\deg}$\,4655      & 41140 $\pm$  236 &  5.724 $\pm$ 0.125 &  1.530 $\pm$ 0.116\\
PG\,2352+181      & 46926 $\pm$  515 &  6.139 $\pm$ 0.084 &  0.607 $\pm$ 0.086\\
UVO\,0832-01      & 47213 $\pm$  712 &  6.136 $\pm$ 0.112 &  0.608 $\pm$ 0.117\\
PG\,0314+146      & 47592 $\pm$  717 &  6.293 $\pm$ 0.113 &  0.506 $\pm$ 0.101\\
BD+48${\deg}$\,1777      & 47681 $\pm$  678 &  6.296 $\pm$ 0.106 &  0.548 $\pm$ 0.101\\
PG\,0838+133      & 48024 $\pm$  804 &  6.039 $\pm$ 0.113 &  0.516 $\pm$ 0.109\\
PG\,2213-006      & 48100 $\pm$  834 &  6.144 $\pm$ 0.118 &  0.293 $\pm$ 0.088\\
PG\,0039+135      & 48436 $\pm$  684 &  6.103 $\pm$ 0.097 &  0.646 $\pm$ 0.117\\
PG\,1427+196      & 48963 $\pm$  616 &  5.927 $\pm$ 0.105 &  0.240 $\pm$ 0.063\\
PHL\,540          & 49156 $\pm$  844 &  6.123 $\pm$ 0.114 &  0.388 $\pm$ 0.099\\
UVO\,0904-02      & 50176 $\pm$  804 &  6.262 $\pm$ 0.115 &  0.458 $\pm$ 0.110\\
FBS\,0730+617     & 50416 $\pm$ 1098 &  5.952 $\pm$ 0.145 &  0.432 $\pm$ 0.135\\
PG\,0952+518      & 51059 $\pm$  829 &  6.206 $\pm$ 0.116 &  0.614 $\pm$ 0.145\\
HS\,1707+6121     & 51134 $\pm$  720 &  5.977 $\pm$ 0.101 &  0.384 $\pm$ 0.055\\
Feige26         & 62434 $\pm$ 1773 &  5.471 $\pm$ 0.126 &  0.339 $\pm$ 0.136\\
PG\,0823+546      & 72293 $\pm$ 2842 &  6.053 $\pm$ 0.137 & -0.071 $\pm$ 0.100\\
\hline
\end{tabular}
\label{table:nonlin}
\end{table}

\section{Discussion} The monitoring of field sdO stars to uncover pulsations has been an ongoing effort over the past decade and has resulted in posing more questions than have been answered. The Fourier analysis for the $\omega$ Cen sdO pulsators showed pulsation peaks with amplitudes from 0.5\% to 2.7\% of the mean stellar brightness. \citet{2006MNRAS.371.1497W} identified pulsations down to an amplitude limit of about 0.25\%, plus a few additional potential peaks that were even fainter. We did not detect any peaks above 0.08\% in any of our 36 targets. Even though we observed only 8 H-rich sdO stars strictly comparable to the $\omega$ Cen pulsators, i.e. whose temperatures place them in or near the $\sim$47,000--56,000~K instability strip, this is still quite puzzling. Comparing our data set with the data set of \citet{2011ApJ...737L..27R}: 123 EHB stars were selected from the ULTRACAM field, and 35 candidates had photometry good enough to detect low amplitude pulsations down to 0.5\% in the u' band (Suzanna Randall, private communication). Excluding the (pre-selected) pulsators, 5 out of their 55 spectroscopic targets (which don't all overlap with the photometry) are H-rich and lie near or in the $\omega$ Cen instability strip. We would therefore expect less than 4 of their well-monitored 35 EHB stars to be both H-rich and in/near the instability strip, yet in fact 3 of those stars were found to pulsate. None of our 8 comparable H-rich field stars were observed to pulsate down to a much lower detection limit. This does not even take into account the fact that there might be undiscovered pulsators in $\omega$ Cen with amplitudes below the detection limit of \citet{2011ApJ...737L..27R}.\\
\indent If it is true that there are no field counterparts to the sdO pulsators in $\omega$ Cen, this surely must be
telling us something about different formation channels for these stars. The one-of-a-kind pulsator, J160043+074802.9, has an effective temperature much hotter than the tightly clustered $\omega$ Cen pulsators and a significantly higher helium abundance, so it is not at all clear whether it is related. Several theories have been proposed which mainly have to do with formation channels of these low mass, evolved, pulsating stars. Are there different formation channels for GC pulsating sdO's than for field sdO's? One would expect that in a higher density region, such as a GC, there would be a higher probability of dynamical interactions. This could potentially lead to a greater chance of the formation scenario where two He white dwarfs (He-WD) merge to become a He-sdO \citep{2012MNRAS.419..452Z}. Field sdO stars in the galactic disk must have formed in much lower density conditions and are therefore considerably less likely to have been produced by a collisional scenario.

\bibliography{author}

\begin{thebibliography}{}
\expandafter\ifx\csname natexlab\endcsname\relax\def\natexlab#1{#1}\fi
\expandafter\ifx\csname url\endcsname\relax
  \def\url#1{\texttt{#1}}\fi
\expandafter\ifx\csname urlprefix\endcsname\relax\def\urlprefix{URL }\fi
\providecommand{\eprint}[2][]{\url{#2}}

\bibitem[{{Latour} et~al.(2011){Latour}, {Fontaine}, {Brassard}, {Green},
  {Chayer}, \& {Randall}}]{2011ApJ...733..100L}
{Latour}, M., {Fontaine}, G., {Brassard}, P., {Green}, E.~M., {Chayer}, P., \&
  {Randall}, S.~K. 2011, \apj, 733, 100

\bibitem[{{Randall} et~al.(2011){Randall}, {Calamida}, {Fontaine}, {Bono}, \&
  {Brassard}}]{2011ApJ...737L..27R}
{Randall}, S.~K., {Calamida}, A., {Fontaine}, G., {Bono}, G., \& {Brassard}, P.
  2011, \apjl, 737, L27

\bibitem[{{Rodr{\'{\i}}guez-L{\'o}pez}
  et~al.(2007){Rodr{\'{\i}}guez-L{\'o}pez}, {Ulla}, \&
  {Garrido}}]{2007MNRAS.379.1123R}
{Rodr{\'{\i}}guez-L{\'o}pez}, C., {Ulla}, A., \& {Garrido}, R. 2007, \mnras,
  379, 1123

\bibitem[{{Woudt} et~al.(2006){Woudt}, {Kilkenny}, {Zietsman}, {Warner},
  {Loaring}, {Copley}, {Kniazev}, {V{\"a}is{\"a}nen}, {Still}, {Stobie},
  {Burgh}, {Nordsieck}, {Percival}, {O'Donoghue}, \&
  {Buckley}}]{2006MNRAS.371.1497W}
{Woudt}, P.~A., {Kilkenny}, D., {Zietsman}, E., {Warner}, B., {Loaring}, N.~S.,
  {Copley}, C., {Kniazev}, A., {V{\"a}is{\"a}nen}, P., {Still}, M., {Stobie},
  R.~S., {Burgh}, E.~B., {Nordsieck}, K.~H., {Percival}, J.~W., {O'Donoghue},
  D., \& {Buckley}, D.~A.~H. 2006, \mnras, 371, 1497.
  \eprint{arXiv:astro-ph/0607171}

\bibitem[{{Zhang} \& {Jeffery}(2012)}]{2012MNRAS.419..452Z}
{Zhang}, X., \& {Jeffery}, C.~S. 2012, \mnras, 419, 452

\end{thebibliography}
\acknowledgements CBJ would like to thank EG for her continuing support and mentorship, without her, this proceeding and the followup paper(s) would not be possible. EG and CBJ would also like to thank Suzanna K. Randall for sharing some of her most recent results in advance of publication and the numerous emails that lent support to the statistical analysis.

\end{document}